\documentclass[11pt]{article} 
\usepackage[utf8]{inputenc}
\usepackage{multirow}
\usepackage{amsfonts}
\usepackage{amsmath}
\usepackage{amssymb}
\usepackage{amsthm}
\usepackage{caption}
\usepackage[dvipsnames]{xcolor}
    \definecolor{darkgreen}{rgb}{0,0.5,0}
    \definecolor{darkblue}{rgb}{0,0,0.6}
    \definecolor{purple}{rgb}{0.4,.2,0.7}
\usepackage[margin = 2.5cm]{geometry}
\usepackage{graphicx}
\usepackage[hyperfootnotes = false, colorlinks = true, linkcolor = darkblue, citecolor = purple]{hyperref}
\usepackage{subcaption}
\usepackage{ytableau}

\usepackage{comment}

\newcommand{\be}{\begin{equation}}
\newcommand{\ee}{\end{equation}}
\newcommand{\bea}{\begin{eqnarray}}
\newcommand{\eea}{\end{eqnarray}}
\def\la{\label}
\def\nref#1{(\ref{#1})}
\def\half{{1 \over 2 }}
\def\slashed#1{{\not\!#1}}

\def\breta{{\bar \eta}}

\begin{document}

\thispagestyle{empty}
\begin{center}
    ~\vspace{5mm}

  \vskip 2cm 
  
   {\LARGE \bf 
       Three Point Amplitudes in Matrix Theory 
   }

   \vspace{0.5in}
     
   {\bf  Aidan  Herderschee and Juan Maldacena 
   }

    \vspace{0.5in}

  Institute for Advanced Study,  Princeton, NJ 08540, USA

    \vspace{0.5in}

    \vspace{0.5in}
    

\end{center}

\vspace{0.5in}

\begin{abstract}
 
  \end{abstract}
 
 We compute the three graviton amplitude in the Banks-Fischler-Shenker-Susskind matrix model for M-theory. Even though the three point amplitude is determined by super Poincare invariance in eleven dimensional M-theory, it requires a non-trivial computation in the matrix model. We consider a configuration where all three gravitons carry non-zero longitudinal momentum. To simplify the problem, we compactify one additional dimension and relate the amplitude to a supersymmetric index computation. We find agreement with the expected answer even at finite values of $N$.

\vspace{1in}

\pagebreak

\setcounter{tocdepth}{3}
{\hypersetup{linkcolor=black}\tableofcontents}

\section{Introduction and motivation }

The BFSS matrix model can be used to compute flat space scattering amplitudes in M-theory \cite{deWit:1988wri,Banks:1996vh,Susskind:1997cw,Seiberg:1997ad,Sen:1997we,Polchinski:1999br}. In this paper, we compute the simplest such amplitude, the three point amplitude, obtaining the expected Lorentz invariant answer. In fact, Lorentz invariance of the eleven dimensional bulk theory uniquely fixes the three point amplitude. However, this is insufficient from the perspective of the matrix model, because eleven dimensional Lorentz symmetry is not present in the matrix model at finite $N$. 

In general, it is difficult to compute amplitudes from the matrix model, because the prescription involves a low energy computation that is strongly coupled. Furthermore, the ground state wavefunctions to be scattered are not known, except in particular limits \cite{Yi:1997eg,Sethi:1997pa,Moore:1998et,Konechny:1998vc,Porrati:1997ej,Sethi:2000zf,Lin:2014wka}. Even though the three point amplitude is fixed by symmetry in M-theory, in the matrix model it corresponds to a highly non-trivial process where two strongly coupled clouds of D0 branes merge into a single cloud of D0 branes. This process seems to be inaccessible by the background field and worldline approximations previously applied at four point where the clouds of D0 branes are always well separated and their internal structure is largely irrelevant \cite{Douglas:1996yp,Becker:1997wh,Becker:1997xw,Polchinski:1997pz,Plefka:1998ed}.

The crucial step in our calculation involves relating the three point amplitude in the matrix model to an index. The index is simply a number and therefore invariant under changes of the coupling. For this relation, it is crucial that the kinematics for which the on-shell three-point amplitude is non-vanishing preserve 1/4 of the supersymmetries. Notably, this configuration involves making one of the momenta imaginary.\footnote{One could also say that we are going to $(2,9)$ signature, but we prefer to view it as an analytic continuation from the usual signature in this paper.} Relating the amplitude to the index involves two steps:
\begin{enumerate}
\item Compactifying one of the transverse dimensions so that the Matrix model becomes a 1+1 gauge theory or matrix field theory \cite{Taylor:1996ik,Dijkgraaf:1997vv}.
\item Introducing transverse M2 branes that are localized along the light-cone direction and are separated along some of the transverse dimensions, including the (Euclidean) time direction. The three separated M2 branes can be viewed as sourcing the three gravitons being scattered. Introducing these M2 branes in the bulk theory amounts to imposing particular boundary conditions in the 1+1D matrix model. 
\end{enumerate}

\begin{figure}[t!]
    \begin{center}
    \includegraphics[scale=0.4]{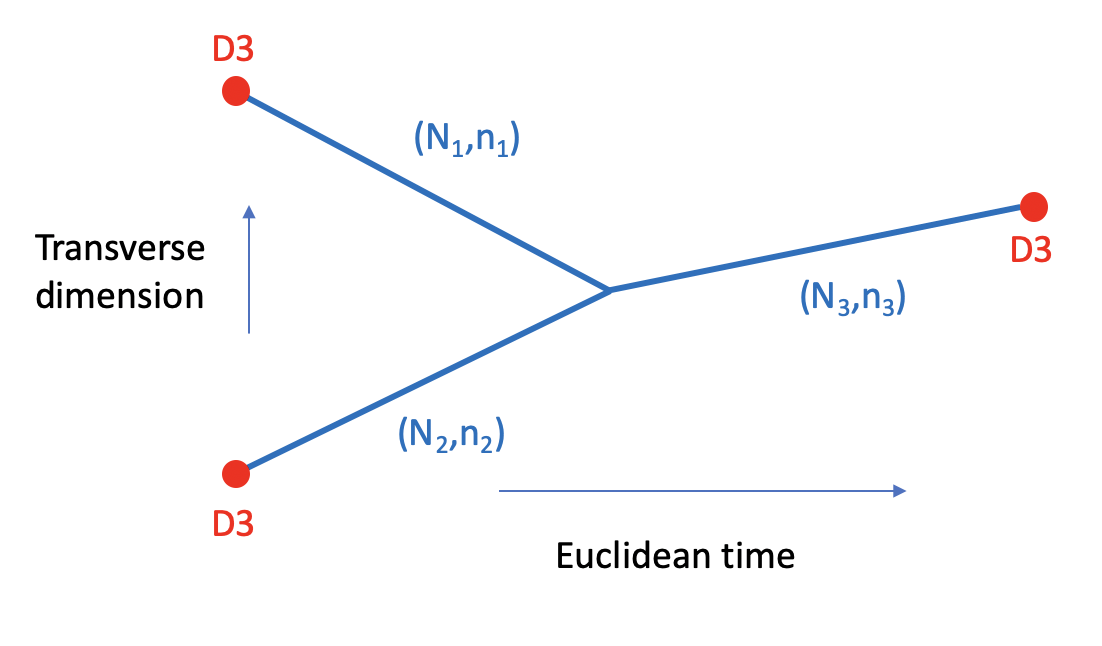}
    \end{center}
    \caption{This is a sketch of the main idea for this paper. The blue lines represent  the scattering amplitude configuration in Euclidean time. $N_i$ and $n_i$ indicate the quantized momenta in the longitudinal null direction and the transverse compact dimension respectively.  The vertical direction is a non-compact transverse dimension. $N_i$ and $n_i$ can also be viewed as the number of D1 branes and fundamental strings that are forming bound state. As we explain below, the three point amplitude is then related to a string network stretching between three separated D3 branes.    }
    \label{StringNetwork}
\end{figure}

Under T-duality, in the 1+1 matrix string description, these separated M2 branes become boundary conditions similar to those imposed by D1 branes ending on D3 branes. The bulk scattering configuration can therefore be identified with a three string network \cite{Bergman:1997yw,Bergman:1998gs,Schwarz:1996bh,Aharony:1997ju,Aharony:1997bh,Dasgupta:1997pu,Sen:1997xi,Kol:1998zb} ending on three separated D3 branes, see figure \ref{StringNetwork}. The number of states corresponding to this string network configuration was computed by Sen in \cite{Banerjee:2008pu,Sen:2008ht,Sen:2012hv}. We can view the problem in an ``open string channel'', where it has an index interpretation, and also in a ``closed string channel'', where it has an amplitude interpretation. The connection between the two is worked out in detail to reproduce that the amplitude depends on continuous parameters, as opposed to the index, which is just a number.

In the limit that $R_9/l_p \to 0$, the matrix model reduces to weakly coupled IIA string theory. It was argued that the matrix string gives rise to string perturbation theory in \cite{Motl:1997th,Dijkgraaf:1997vv}, which reproduces the gravity three point function, see also \cite{Giddings:1998yd,Bonelli:1998wx,Bonelli:1999qa}. 
In this paper, we argue that this is indeed the case for all values of the coupling, or all values of $R_9/l_p$, enabling us to access the eleven dimensional limit $R_9 \to \infty$.

This paper is organized as follows. 
In section \ref{sec:twothreeampmatrix}, we write the two and three point amplitudes as determined by supersymmetry and Lorentz invariance. We rewrite the amplitudes in a few different ways so as to take them to a form that is most natural for the matrix model computation. 
In section \ref{sec:bfssmodel}, we recall a few facts about the matrix model, introduce the setup for our computation, and compute the three point amplitude. Finally, we present some conclusions and discussion. 

\section{Two and three point amplitudes in M-theory } \label{sec:twothreeampmatrix}

M-theory contains massless excitations that are in a short multiplet of the supersymmetry algebra. In fact, a massless particle preserves half of the supersymmetries, the ones that obey $\slashed{p} \epsilon =0$. 

 In order to describe the kinematic configuration of the three point amplitude, it is convenient to select four of the eleven dimensions and introduce the following coordinates 
 \be 
 \begin{split} \la{MetDefin}
 &d s^2 = - 2 dx^+dx^- + 2 dz d\bar z ~, \\
 &~~\textrm{where} ~~~\sqrt{2} z = x^9 + i x^8 , ~~\sqrt{2}\bar z = x^9 - i x^8, ~~\sqrt{2}x^{\pm}=x^{0}\pm x^{10} \ .
 \end{split}
 \ee  
 We can now choose a configuration with momenta that are purely in the $p_-$ and $p_z$ directions, with $p_+ = p_{\bar z } =0$ for all particles. This is possible only if the momenta in the direction $p_8$  are purely imaginary. For example, one possible configuration is
 \be \la{MomExp}
 p^i_\mu =(p_+,p_-,p_z , p_{\bar z } ) ~,~~~~ p^1_\mu = ( 0,p^1_-, q_z , 0 ) ~,~~~~p^2_\mu =(0,p^2_-, -q_z, 0) ~,~~~~p^3_\mu =- (0,p^1_ - + p^2_-, 0,0)   
\ee 
which clearly obeys momentum conservation. In addition, imposing ${\slashed{p}}^i \epsilon =0$ for the three particles $i=1,2,3$ implies that $\Gamma^- \epsilon = \Gamma^z \epsilon =0 $, which are compatible conditions that select 1/4 of the 32 original supersymmetries.\footnote{This means that out of the 32 supersymmetric Ward identities \cite{Grisaru:1977px} that we have for generic on-shell amplitudes only 24  linearly independent ones remain for the three point amplitude. }

In order to write the scattering amplitudes, it is convenient to use an efficient description for the whole supermultiplet. Since any on-shell three point amplitude configuration can be related by Lorentz symmetry to one where the momenta are in four of the dimensions, as in \nref{MomExp}, we will use four dimensional notation for the supersymmetries. This is the same notation that is used in describing ${\cal N}=8$ supergravity in four dimensions. 
In that case, it is convenient to use spinor helicity variables to write the momenta as $p_{\alpha \dot \beta } = \lambda_{\alpha } \bar \lambda_{\dot \beta }$. 
The supercharges obey an algebra of the form 
\be 
\la{SusyAlg} \{  Q_{\alpha I } , \bar Q_{\dot \beta }^J \} = \lambda_{\alpha } \bar \lambda_{\dot \beta }\delta_I^J
\ee 
 and we can generate the whole multiplet by starting from the positive helicity graviton and writing \cite{Nair:1988bq,Witten:2003nn,Bianchi:2008pu,Arkani-Hamed:2008owk}
\be 
|\bar \eta \rangle = e^{ b^{\dagger I } \bar \eta_I } |+ \rangle ~,~~~~~ {\rm with }~~~~~   b^{\dagger I } \equiv \bar w^{\dot \beta} \bar Q_{\dot \beta}^{\, I  }    
\la{SMchi}
\ee 
where $\bar w$ and $w$ are spinors obeying\footnote{ Note that $[\bar w, \bar \lambda]$ denotes a ``square bracket'' defined as in \nref{omegaDef} and {\it not} a commutator.}  
\be \la{omegaDef} 
[ \bar w , \bar \lambda ] \equiv {\bar w}^{\dot \beta } \bar \lambda_{\dot \beta } = 1 ~,~~~~~~~~~ \langle w , \lambda \rangle = w^\alpha \lambda_\alpha =1 \ .
\ee 
In \nref{SMchi}  $I$ runs over eight values. This implies that $b^{\dagger I}$ in \nref{SMchi} and $b_I \equiv w^\alpha Q_{\alpha I} $ are canonically normalized fermion creation and annihilation operators obeying $\{ b_{I} , b^{\dagger I' } \} = \delta_I^{I'}$ thanks to \nref{SusyAlg} and \nref{omegaDef}.

Using this notation, the two point amplitude is 
\be \la{RelTwo}
{\cal A}_2 = - 2 p_- (2 \pi)^{10}\delta(p_- + p'_-) \delta^9( \vec p + \vec p') \delta^8( \bar \eta_I + \bar {\eta'}_I ) 
  \ee 
  (note that there is no $\delta(p_+ + p_+')$ term). 
  This sets the normalization of the states.

For the three point amplitude, we will consider a configuration where all $\bar \lambda^i$ are proportional to each other\footnote{The other amplitude  is similar up to exchanging the bar and unbarred quantities \cite{Arkani-Hamed:2008bsc}, and we will not discuss it any further in this paper.   } 
\be \la{Athr}
{\cal A}_3
=    \sqrt{16 \pi G_{N,11}} \sqrt{2} (2\pi)^{11}  \delta(\sum p_-^i) \delta(\sum p_+^i) \delta^9(\sum \vec p^{\, i}) \delta^{16} ( \lambda_{\alpha}^i \bar \eta_I^i ) { 1 \over \left( \langle 1,2\rangle \langle 2,3 \rangle \langle 1,3 \rangle \right)^2 }
\ee 
%
This form of the amplitude is completely determined by supersymmetry and Lorentz invariance. In particular, supersymmetry fixes the sixteen delta functions in \nref{Athr}. This is, of course, the amplitude in the gravity theory, but it is also the amplitude in the full M-theory because symmetries fully fix it up to an overall constant. Here we fixed the overall constant by computing it in gravity \cite{DeWitt:1967ub}. 

\subsection{Amplitudes in gravity with compact directions } 

For our problem, we will be interested in compactifying the $x^- $ direction. In addition, we also find it convenient to compactify in the $x^9$ direction. The momenta are then \footnote{Note that $p_\pm$ with the indices down are negative semidefinite for physical ingoing particles.} 
\be 
-p_-^i = { N_i \over R_-} ~,~~~~~~p_9 = -i p_8 = { n_i \over R_9 } ~,~~~~~ p_z = \sqrt{2}  { n \over R_9} ~,~~~~{\rm with} ~~x^- \sim x^- + 2\pi R_-~,~~~~x^9 \sim x^9 + 2 \pi R_9 
\ee 
$p_9$ is quantized and  $p_8$ is fixed to this value by the kinematics of the three point amplitude even though $x^8$ is non-compact. 
These momenta correspond to the spinor helicity variables 
%
\bea 
\lambda_i  &=&  \left( \begin{array}{c} -\sqrt{ N_i \over R_-}  \cr 
 \sqrt{2}{ n_i \over R_9 } \sqrt{R_- \over N_i}   \end{array} \right) ~,~~~~~ 	\bar \lambda_i =   \left( \begin{array}{c}  \sqrt{ N_i \over R_-}\cr 
 0  \end{array} \right)  ~~~~i=1,2
 \cr 
 \lambda_3  &=&  \left( \begin{array}{c} -\sqrt{  N_3 \over R_-}  \cr 
  \sqrt{2}{ n_3 \over R_9 } \sqrt{R_- \over  N_3}   \end{array} \right) ~,~~~~ 	\bar \lambda_3 =  - \left( \begin{array}{c}  \sqrt{   N_3 \over R_-}\cr 
 0  \end{array} \right)	~,~~~ N_3 = N_1 + N_2 ~,~~~n_3 = n_1 + n_2 \la{ThrT}
\eea 
We have also chosen the integers so that momentum conservation is obeyed. These give momenta similar to \nref{MomExp}, where $p_+^i =0 = p_{\bar z}^i$. We are writing the momenta in the all ingoing notation and that is the reason for the extra minus sign in \nref{ThrT}. We are viewing particles 1 and 2 as ingoing and 3 outgoing.  

After these choices, the amplitude becomes, with $16\pi G_{N,11} = { (2 \pi)^8  } l_p^9$,   \bea
 {\cal A}_2 & = &2 N R_9  (2\pi)^{10} \delta^8(\vec p + \vec p') \delta^8( \bar \eta_I + \breta'_{I}) ~,~~~~~~~~~N= N'~,~~~n=n'
\\
\la{AthrInt}
{\cal A}_3
 &=  &    (2 \pi)^4  (2\pi)^{11} 2 \sqrt{2}  l_p^{9/2} { R_- \over R_9 }    { ( N_1 N_2 N_3)^2 \over (n_1 N_2 - n_2 N_1)^6   }  \delta(\sum p_+^i) \delta^8(\sum \vec p^{\, i}) \delta^{8}(\sum  \sqrt{N_i} \bar \eta^i_I) \delta^8( \sum  { n_i \over \sqrt{N_i } }\bar \eta^i_I  )   ~~~~~~~~~~
 \eea 
 where we have used $\delta(n/R) \sim R \delta_n$ and conservation conditions \nref{ThrT} for the integers. These remove the $p_-$ and $p_9$ delta functions. The $p_8$ delta function is still present in the above expression.

 \subsection{Rewriting the amplitudes }  
  
 In order to compare to the matrix model, it is natural to normalize the states using the cannonical normalization than the relativistic one \nref{RelTwo}:
 \be \la{NorNew}
 {\cal A}_2  = \langle p | p'\rangle ~,~~~~~~~{\cal A}_2^c = ~^c\langle p | p'\rangle^c ~,~~~~~|p\rangle^c = { |p \rangle \over 2\pi \sqrt{2  N R_9 } } \ .
 \ee 
We will also switch the variables $\breta_1, ~\breta_2$ to ``center of mass'' and ``relative'' variables 
\be \la{EtaCMr}
  \breta^1_I =   \breta^{cm}_I \sqrt{N_1 \over N_3}    + \breta^r_I \sqrt{ N_2 \over N_3} ~,~~~~~~~\breta^2_I =   \breta^{cm}_I \sqrt{N_2 \over N_3}    - \breta^r_I \sqrt{ N_1 \over N_3} \ .
  \ee 
  At this point, this change of variables looks arbitrary. However, these new variables are convenient for the matrix model computation. Applying this redefinition, \nref{EtaCMr},  with the canonical normalization \nref{NorNew}, the amplitudes become  
 \bea
 {\cal A}_2^c & = &  (2\pi)^8\  \delta^8(\vec p + \vec p') \delta^8( \bar \eta_I + \breta'_{I}) ~,~~~~~~~~~N= N'~,~~~n=n' \ ,
\\
\la{AthrIntc}
{\cal A}_3^c
 &= &     { (2 \pi)^4 } { 1 \over 2 \pi  }  l_p^{9/2} { R_- \over R_9^{5/2}  }  {    (n_1 N_2 - n_2 N_1)^2 N_3^2 \over ( N_1 N_2  )^2  \sqrt{N_1 N_2 N_3 } }  (2\pi)^9 \delta(\sum p_+^i) \delta^8(\sum \vec p^{\, i}) \delta^{8}(\breta^{cm}_I + \breta^3_I) \delta^8(  \breta^r_I)  ~~~~~~~~   \ .
 \eea 
 Equation \nref{AthrIntc} is simply a rewriting of the original amplitude in non-relativistic normalization.
 
We will do one more change of variables. It will be more convenient for us to express the supermultiplet  in terms of supercharges acting on a different ``vacuum'' than the one in \nref{SMchi}. We want to pick a vacuum that is annihilated by a supercharge obeying 
\be \la{SusyM2}
 \Gamma^{-+ 456789} \epsilon = \epsilon \ .
 \ee 
 This selects the supersymmetries that are preserved by an M2 brane along the directions 123. 
It is useful to  split the eight supercharges we had in \nref{SMchi} into two groups with opposite eigenvalues under the projector $\Gamma^{-+ 456789} \epsilon = \pm  \epsilon$. 
 . We will denote them in terms of the indices $J$ and $K$, each of them going from one to four (while the original index $I$ went from one to eight). 
 In other words, $\breta_I \to (\breta_J , ~\breta_K)$. After making this replacement, we then ``Fourier transform'' the $\breta_K $ to $\eta^K$. In other words, we define a ``new" amplitude
 \be 
 \hat {\cal A}_n(\breta^i_J,\eta^{iK}) = \int \prod_i d^4\breta^i_K e^{ \eta^{iK} \breta_K^i } {\cal A}_n(\breta^i_J,\breta^i_K) 
 \ee 
which is an amplitude of the supersymmetric coherent states
\be \la{NewVac}
|\breta_J, \eta^K \rangle = e^{  [ \bar{w} , \bar Q^J] \bar \eta_J } e^{ \langle w , Q_K \rangle \eta^K } | 0 \rangle 
\ee 
where $|0\rangle $ is a particular state with spin zero in four of the dimensions, namely $|0\rangle = \prod_{K=1}^4 [ \bar w , \bar Q^K] |+ \rangle$, with $|+\rangle$ as in \nref{SMchi}.\footnote{This is the supergravity analog of the non-chiral superspace studied in \cite{Huang:2011um} for super Yang-Mills.} 

 Defining the $\eta^K$ version of the center of mass and relative variables as in \nref{EtaCMr}, the amplitudes can be rewritten as
  \bea \la{TwoF}
 \hat {\cal A}_2^c & = & (2\pi)^8  \delta^8(\vec p + \vec p') \delta^4( \bar \eta_J + \breta'_{J})\delta^4(   \eta^K - {\eta'}^{K}) ~, ~~~~~~N=N' ~,~~~~~n=n' \\
\la{ThreeF}
\hat {\cal A}_3^c
 &= &    (2\pi)^8  \delta^8(\sum \vec p^{\, i}) \delta^4( \bar \eta^{cm}_J + \breta^3_{J})\delta^4(   \eta^{cm K} - \eta^{3K}) \times 
 \cr 
 &~& \times   { (2\pi)^4} { 1 \over 2 \pi   } l_p^{9/2} { R_- \over R_9^{5/2}  }  {    (n_1 N_2 - n_2 N_1)^2 N_3^2 \over ( N_1 N_2  )^2  \sqrt{N_1 N_2 N_3 } }  2\pi \delta(\sum p_+^i) 
      \delta^4(  \breta^r_J)     \ .
 \eea 
Note that the first line in \nref{ThreeF} has a form similar to the two point function \nref{TwoF}, a fact which will be exploited later.

Let us emphasize that we have done nothing nontrivial yet. The amplitudes \nref{TwoF} and \nref{ThreeF} are trivial rewritings of the amplitudes \nref{RelTwo} and \nref{Athr} in terms of convenient variables. The goal of Section \ref{sec:bfssmodel} is to derive \nref{TwoF} and \nref{ThreeF} from a matrix model computation.

 \section{The BFSS Matrix model } \label{sec:bfssmodel}

 We first review the BFSS model and emphasize some aspects that will be important in later computations. The BFSS matrix model is defined by 
 \be \la{LagBFSS}
 S = { 1 \over R_- } \int dt \textrm{Tr}\left[ \half \sum_{a=1}^9 ( D_t X^a)^2 + { 1 \over 4 } { R_-^2 \over ( 2 \pi)^2 l_p^6 } \sum_{a,b=1}^9 [ X^a , X^b]^2 + {\rm  fermions } \right] 
 \ee 
 where $t = x^+$,  and $X^a$ are normalized so that they are matrix versions of the position coordinates, with dimensions of length. The matrices are $N\times N$ in the adjoint of the $U(N)$ gauge group  \cite{Banks:1996vh,Taylor:1997dy}. The fermions are in the {\bf 16} of SO(9).  We also used that $G_{N,11} = 16 \pi^7 l_p^9$. 
   
  The $U(1)$ part of the theory decouples and describes the center of mass degrees of freedom, which consist of    nine positions and 16 Majorana fermions. 
  These fermions are the Goldstone fermions of the  16 supercharges that are not preserved by a massless superparticle with null momentum.  
   The physics of these nine variables and 16 fermions reproduces what we expect for a superparticle in light cone gauge. In particular, they give rise to the two point amplitude \nref{TwoF} and the ``center of mass'' part of the three point amplitude. We will discuss this in more detail below. 
   The $SU(N)$ degrees of freedom are believed to give rise to a zero energy bound state at threshold \cite{Witten:1995im} as well as a continuum.  The continuum is related to configurations where the matrices develop large diagonal vacuum expectation values that break $U(N) \to \prod_{i=1}^\ell  U(N_i)$ which are then viewed as a scattering state of $\ell$ gravitons with null momentum $-p_-^i = N_i /R_-$.  Each of these $U(N_i)$ factors has a $U(1)_i$ part and an $SU(N_i)$ part.   When we define the asymptotic scattering states we focus on the zero energy bound state in the $SU(N_i)$ parts. Then the $U(1)_i$ parts describe massless superparticles.

   The 32 original supersymmetries are still realized in this theory: 16 of them linearly and 16 non-linearly. The latter are realized as shifts of the 16 Majorana fermions of the $U(1)$ part of the theory. The quantization of these sixteen fermions in each $U(1)$ fills out the $2^8$ states of a massless superparticle in eleven dimensions. We can use the same method as in \nref{SMchi} to label the asymptotic states in terms of Grassmann variables. This means that we can define a matrix model amplitude, in canonical normalization,  in terms of scattering states 
  $
   {\cal A}_{MM}^c 
   $.  
   The BFSS conjecture \cite{Banks:1996vh} states that the large $N$ limit of these amplitudes, keeping $-p^i_- = N_i/R_-$ fixed, precisely matches the M-theory amplitudes with $x^-$ non-compact.  
   
   The matrix model does not realize the full super-Poincare symmetry, but it does realize a subgroup of these symmetries. This subgroup is constraining enough to fix the full polarization dependence of the three point amplitude. To see this, first note that, for the three point amplitude,  we can use rotation symmetry, momentum conservation and Galilean symmetry (which is a part of the original Lorentz group that commutes with $x^-$ translations) to transform to a momentum configuration that lies in just four of the eleven dimensions. This justifies our use of four dimensional notation.  This is completely parallel to our discussion of the M-theory amplitude. The only new point is that we do {\it not} need to use the broken boosts that involve the $x^-$ coordinate, the boosts  that would change $N_i$. 
   
  The next observation is that the  algebra of the supercharges is not modified relative to the non-compact situation.  In other words, the supercharge algebra acting on the external states is the same as in \nref{SusyAlg}. All the various polarization states can be obtained by the action of the same operators discussed in \nref{SMchi} and \nref{omegaDef}. In the matrix model, the $b_I$ and $b^{\dagger I}$ are the fermionic creation and annihilation operators for each $U(1)_i$ factor \cite{Plefka:1997xq}. Once we combine the various polarization states into a field as in \nref{SMchi}, we can then view the supercharges as acting as 
   \be \la{SusyExp}
    Q_{\alpha I} = \sum_i \lambda^i_\alpha \bar \eta_{I}^i ~,~~~~~~    \bar Q_{\dot \beta }^I = \sum_i \bar \lambda^i_{\dot \beta } \partial_{\bar \eta^{i}_{I} } \ .
    \ee 
    For three point kinematics, equation \nref{SusyExp} gives us the action of only half of the $\bar Q$ supersymmetries. However, this is sufficient because the three point function is actually invariant under the other half. 
   We now try to write the most general Grassman expression for the   three point function that will be invariant under \nref{SusyExp}:
   \be
   \delta^{16} ( \sum_i  \lambda^i_\alpha \bar \eta_{I}^i ) \la{DelCon} \ .
   \ee 
    In principle, equation \nref{DelCon} could be multiplied by an extra function of the $\bar \eta^i_I$. However, for the three point function, there is no function we can add that is non-zero on the support of the delta function \nref{DelCon} which is invariant under the $  \bar Q  $ supersymmetries in \nref{DelCon}. Of course, \nref{DelCon} is invariant due to momentum conservation.  This means that \nref{DelCon} is the only combination that is invariant under supersymmetry. This determines completely the polarization dependence of the three point amplitude. 
    Let us emphasize that despite the fact that we are using four dimensional notation, equation \nref{DelCon} encodes all the polarization data in the full eleven dimensional theory. In other words, the $\bar \eta_I$ variables have spin in the other 7 dimensions. 
   
   The conclusion is that the only unknown part of the three point amplitude is  a function of the $N_i$. These cannot be changed by any of the symmetries that are realized in the matrix model. In other words, the three point amplitude in the matrix model is 
   \nref{Athr}, up to an overall multiplicative function of the $N_i$. 
   We will show that this function matches the gravity answer, including the numerical constant.

  \subsection{The BFSS model after we compactify one spatial dimension} 
   
   We will not attack the problem of computing the three point amplitude directly. We  first consider a slightly different problem which leads to a couple of simplifications. In particular, we compactify one of the spatial dimensions, $x^9 \to x^9 + 2 \pi R_9$. The matrix model is replaced by a 1 +1 dimensional matrix gauge theory \cite{Taylor:1996ik,Banks:1996vh,Dijkgraaf:1997vv}  with the Lagrangian 
 \bea \la{MStr}
    S  
   &=& { 1 \over R_{-} (2 \pi \tilde R_9 ) } \int dt \int_0^{2\pi \tilde R_9} d \tilde x_9 \textrm{Tr}\left[ -
  \half (D_\alpha X^a)^2 + { 1 \over 4 } { R_{-}^2 \over (2 \pi)^2  l_p^6 } [ X^a,X^b]^2 - {1 \over 4 } {( 2 \pi  )^2 l_p^6 \over R_{-}^2 } F^2 +{\rm fermions}\right]  ~~~~~~
  \\ 
  &~& {\rm with }~~~~~~~\tilde R_9 =   { l_p^3 \over R_{-} R_9 } 
  \eea  
   where $\tilde x^9$ is T-dual to the original coordinate $x^9$ \cite{Taylor:1996ik,Seiberg:1997ad,Sen:1997we,Giddings:1998yd}.  
  The two dimensional gauge theory has a quantized total electric flux 
  \be 
  n = {\partial L \over \partial F_{01} } = { 2 \pi l_p^6 \over R_-^3 \tilde R_9} \textrm{Tr}[ F_{01}] \ .
  \ee 
  This integer should be viewed as the momentum in the compact $x^9$ direction. Indeed, we can check that the energy of such a state would be given by $E  = -p_+ = \half { n^2 \over R_9^2} { R_- \over N } $, in agreement with $ 2 p_+ p_-= {n^2\over R_9^2} $.

\begin{figure}[t!]
    \begin{center}
    \includegraphics[scale=.3]{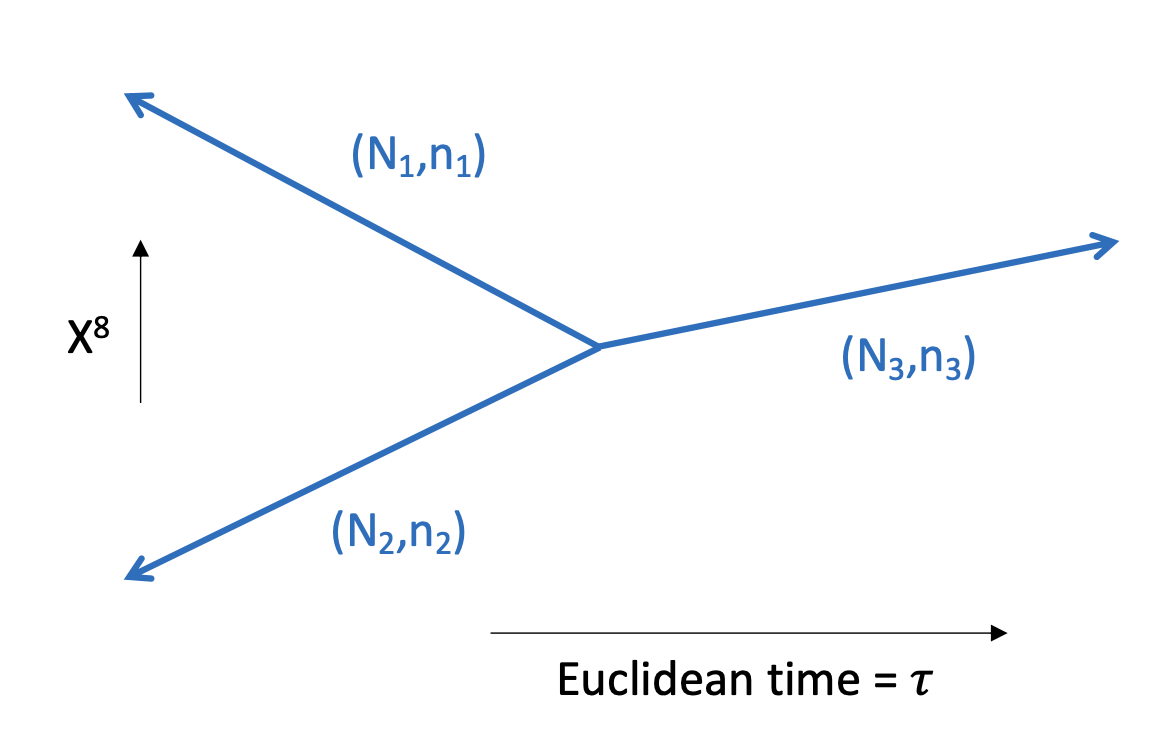}
    \end{center}
    \caption{ The scattering amplitude configuration in Euclidean time. As  $\tau\to -\infty $   the   matrix splits into  two groups $N_1$ and $N_2$. For $\tau \to +\infty$ they  join into a single bound state. For suitable choices of the electric fields,  $n_i$,   the bound states are gapped. We can view this configuration as a (limit of) a simple BPS string network \cite{Sen:1997xi}. The arrows indicate that the D1 branes extend to infinity.}
    \label{AmplitudeEuclidean}
\end{figure}

  The compactification is desirable for two reasons. First, in the case is that the $(N,n)$ are coprime\footnote{``coprime" means that they do not share any non-trivial common factor.} the ground state is gapped \cite{Witten:1995im}. The presence of the gap is useful because it allows the bound state to be more cleanly isolated from the continuum. Second, after recalling that  a real value of $n$ corresponds to real momentum along $x^9$ and imaginary momentum along $x^8$, we can now 
go to  to Euclidean time. Namely, we take  $t=x^+$ in \nref{MStr} imaginary, $t = i \tau$, so that $X^8 \propto n \tau$ with no $i$. Euclidean time naturally leads to a simple picture for the scattering process, which is given in figure \ref{AmplitudeEuclidean}. Notice that euclidean time in the matrix model corresponds to imaginary $x^+$ in the original eleven dimensional description.

 This configuration can be viewed as a limit of an $(N,n)$ string network consisting of branes with $N$ units of D1 brane charge and $n$ units of fundamental string charge. We say it is a limit because we are only considering the field theory   in \nref{MStr}, which is a decoupling limit of the fields on the D1 brane theory. This description as a string network is not strictly necessary but it helps us connect the problem to the work by Sen in \cite{Sen:2012hv}. 
 
 \subsection{Introducing brane sources } 
 
 In order, to relate the computation to an index, we modify the D1 boundary conditions so that the D1 branes terminate at finite Euclidean time. In other words, in the field theory defined by \nref{MStr}, we want to introduce boundary conditions in the Euclidean time direction. These boundary conditions amount to demanding that the D1 branes terminate on D3 branes extended along the directions $123\tilde 9$. 
 We have Dirichlet boundary conditions for the directions tranverse to the D3 branes. For the longitudinal directions, we have a more complicated boundary condition involving a Nahm pole \cite{Nahm:1979yw,Diaconescu:1996rk}
 \be 
  X^A(\tau)  = { L^A \over (\tau - \tau_{\rm bdy}) } + {\rm subleading} ~,~~~~~~~~{\rm a  } ~~\tau \to  \tau_{\rm bdy}~,~~~~~~~~~~A =1,2,3
  \ee 
  where $L^A$ are $N_3 \times N_3 $ matrices obeying the $SU(2)$ commutation relations. For the right boundary we impose that we have a single irreducible representation of $SU(2)$ with dimension $N_3$ or spin $j$  with $2j+1 = N_3$ \cite{Diaconescu:1996rk}. 
  On the left side, the $L^A$ are given by two irreducible representations, one with dimension $N_1$ and one with dimension $N_2$, each forming a sub-block of the matrix $L^A$. In addition, for the $U(1)_i$ part of all the $U(N_i)$, we have Neumann boundary conditions along the longitudinal directions (the directions 1,2,3). For $X^8$, we have the Dirichlet boundary conditions 
 \be 
 X^8(\tau_l) = \left( \begin{array}{cc} x^8_1 {\bf 1}_{N_1} & 0 \\ 0 & x^8_2 {\bf 1}_{N_2} \end{array} \right) 
 ~,~~~~~~~~~X^8(\tau_r) = x^8_3 {\bf 1 }_{N_3} 
 \ee 
 where ${\bf 1}_N$ is the identity matrix of dimension $N$. The other boundary conditions are simply $X^a=0$ for $a=4,5,6,7$. In addition, we have a boundary condition for the electric flux. On the right side, at $\tau = \tau_r$ we impose that the total electric flux is $n_3$. On the left side, there are two $U(1)$ electric fields, one involving the identity in the first $N_1$ entries and the second involving the identity in the second $N_2$ entries of the matrix. We demand that these are $n_1 $ and $n_2$ respectively.

 \begin{figure}[h]
    \begin{center}
    \includegraphics[scale=.4]{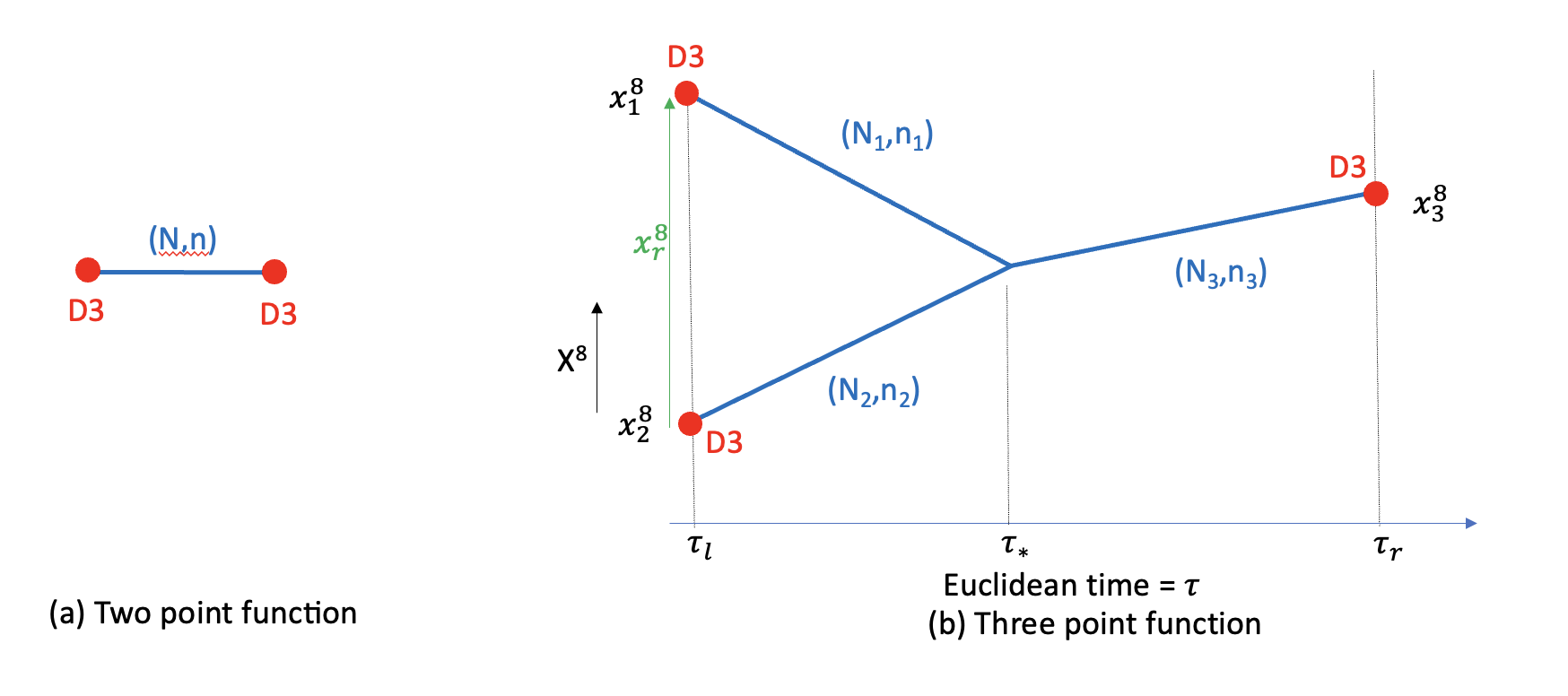}
    \end{center}
    \caption{ The problem with D3 brane sources. We now have the 1+1 field theory with a set of D3 boundary conditions. In (a), we see the problem for the two point function. In (b), for the three point function. We have $N_3 = N_1 + N_2$ and $n_3 = n_1 + n_2$. We have also defined $x_r^8 = x^8_{1} - x^8_2$ to be the relative separation in the $x^8$ direction, which is a boundary condition in the $SU(N)$ part of the 1+1 dimensional theory on the interval. The time $\tau_*$ is well defined in the classical theory, but it is an integration variable in the quantum theory.     }
    \label{D3Amplitude}
\end{figure}

In the original M-theory problem, these boundary conditions amount to adding M2 branes extended along the directions 123. The supersymmetries preserved by these M2 branes obey the condition \nref{SusyM2}. If we consider the gravitons sourced by such M2 branes, we see that they will be annihilated by such supersymmetries. In other words, the graviton has some momentum $p$ and, in addition, the particular polarization state is annihilated by supersymmetries determined by the M2 brane.  This is precisely the vacuum state \nref{NewVac}. In other words, the vacuum, $|0 \rangle$ in \nref{NewVac}, is annihilated by 8 of the supercharges broken by the momentum $p$, the 8 that are preserved by the M2 branes. These are the supercharges $\langle w,Q_J\rangle $ and $ [ \bar w , \bar Q^K]$ where the indices $K$ and $J$ are defined as in the remarks after \nref{SusyM2}.   Here we used the M2 brane language, but this can be easily translated into the field theory language of \nref{MStr}. The reason we chose to write the amplitude as in \nref{ThreeF} is because the $|0\rangle$ ground state is the one   produced by these brane sources. 
  
  Let us emphasize that, even though we talk about branes, the computation we are doing just involves the field theory \nref{MStr} with particular boundary conditions. 
\begin{figure}[h]
    \begin{center}
    \includegraphics[scale=.5]{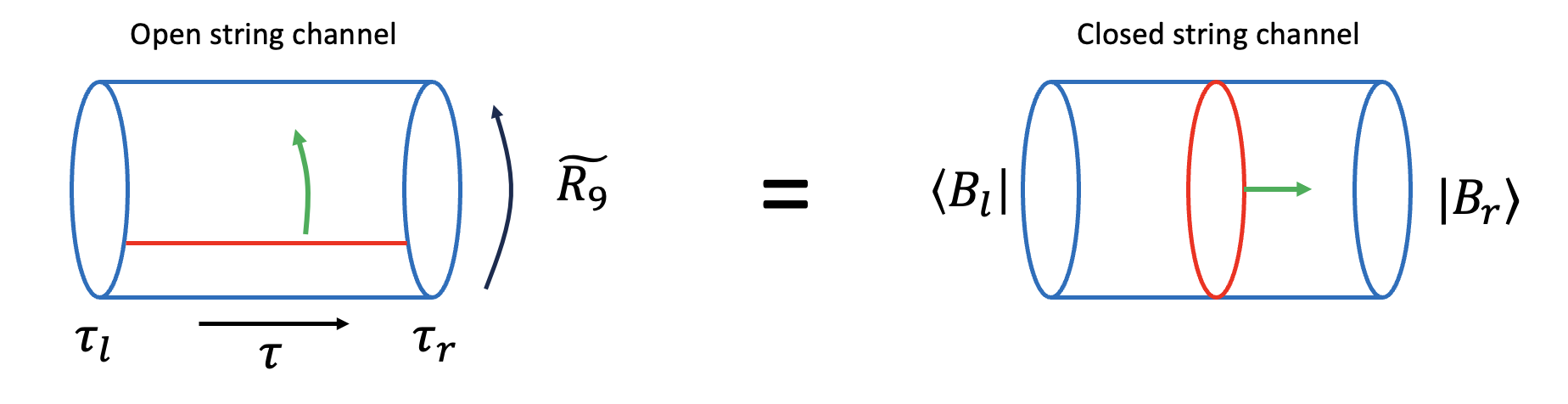}
    \end{center}
    \caption{Two equal views. In the ``open string channel'' we view the theory as defined on a spatial interval and propagating along a circle of Euclidean time. In the ``closed string channel'' we view the theory as defined on a closed circle and propagating between two boundary states. The red line indicates the spatial interval and the green direction the Euclidean time direction.    }
    \label{OpenClosedChannel}
\end{figure}
  This field theory is defined on a cylinder given by the product of a circle along the $\tilde 9$ direction and an interval $[\tau_l, \tau_r] $ along the $\tau$ direction. In this case, we can view the theory in what is called the ``open string channel'' and the ``closed string channel,'' see figure \ref{OpenClosedChannel}. In the first case we view 
  $\tilde x^9$ as the Euclidean time direction and in the second we view $\tau$ as a Euclidean time direction. 
  Note that we are not talking about the conformal field theory of a fundamental string here, but the non-conformal theory described by \nref{MStr}.  
  
  In the open string channel, we are essentially computing an index. The reason is that the fermions obey periodic boundary conditions along the $\tilde x^9$ circle. 
  In the closed string channel, we can view the problem as the overlap between a left and right boundary state up to evolution by the Hamiltonian on the $\tilde R_9$ circle over a period of Euclidean time: 
  \be \label{closedstring}
  Tr\left[ (-1)^F \exp\left( - 2\pi \tilde R_9 H_{\rm open} \right) \right] = \langle B_l | \exp\left( - (\tau_r - \tau_l) H_{\rm closed} \right)  |B_r \rangle \ .
  \ee 
  For the case under study, both the left and right hand sides are zero due to fermion zero modes. However, the equality continues to be true when we insert operators. We will therefore insert an operator that soaks up the fermion zero modes.

  \subsection{The two point function and the boundary states} 
  
  We will first discuss the problem involving the two point function. In this case, we have two D3 branes, see figure \ref{D3Amplitude}(a).  
  It is convenient to split the dynamics into that of the $U(1)$ degree of freedom and the $SU(N)$ part. 
  Once we include a non-zero electric flux $n$, coprime with $N$, we have a gapped system for the $SU(N)$ part \cite{Witten:1995im} and we can forget about it for this problem. In other words, all the contributions for the two point function come from the $U(1)$ part. 
  The configuration with the boundary conditions preserves 8 of the kinematic supercharges (the non-linearly realized ones) and 8 of the linearly realized ones. This means that we have 8 fermion zero modes coming from the kinematic supercharges. 
  This makes the naively defined index vanish. We can get something non-vanishing by computing an index of the form \cite{Sen:2012hv}
  \be \la{Ind8}
  {\cal I}_4 =  \textrm{Tr}\left[(-1)^F { (2J)^4 \over 4!} \right] = 1
  \ee 
  which contains insertions of the angular momentum operator $J$. Each operator $J$ soaks up two fermion zero modes. 

   In the closed string channel, we  have a boundary state of the form 
   \be 
   |B\rangle = {\cal N }  |0 _\parallel\rangle  \int { d^5p \over (2\pi)^5} |p \rangle 
   \ee 
   where ${\cal N}$ is a normalization factor that we will need to fix.  The state $|0_\parallel\rangle $ denotes the zero momentum state in the 123 directions, where we have Neumann boundary conditions. The integral over $p$ enforces the Dirichlet boundary conditions along the other five directions. Notice that we integrate over  five   and not six directions because  one of the directions orthogonal to the D3 brane is the $\tau$ direction.

 The open string channel is a computation of the index with an insertion of $J^4$, which takes the schematic form 
  \be \la{ConC}
  {\cal Z}_{o}\left({(2J)^4 \over 4!}\right) = Tr\left[ (-1)^F { (2J)^4 \over 4!} e^{ - (2\pi \tilde R_9) H_{open} } \right] = Tr\left[ \chi_o^8 e^{ - (2\pi \tilde R_9) H_{open} } \right] = \langle B_l | \chi_o^8 e^{ - \Delta \tau  H_{cl} } |B_{r} \rangle 
  \ee 
  The operators $\chi_o$ are eight different  fermion zero modes canonically normalized  in the open string channel. Each angular momentum generator contains a pair of these fermion zero modes. The factor of 2 in $2J$ was chosen so that $J$ is a product of two canonically normalized fermion zero modes. 
  When we say canonically normalized we mean the following. The fermions on the worldvolume have an action of the form 
  \be 
   S = C \int dt d\tilde x^9 [\psi \partial_t \psi +\cdots ] \sim C\Delta \tau  \int d\tilde x_9 \psi_0 \partial_t \psi_0 = \int d\tilde x_9 \chi_o \partial_t \chi_o ~,~~~~~~ \chi_o = \sqrt{C \Delta \tau  } \psi_0  
   \ee 
   where $ \psi_0$ is the constant part of $\psi$ on the interval and $C$ is a constant that will not matter for us. Inserting $\chi_o$ into \nref{ConC} is equivalent to inserting eight different fermions $\psi$ at any points on the two dimensional cylinder since only the zero mode contributes. 
   On the other hand, when we go to the closed string channel, it is natural to define the canonically normalized fermion, $\chi_{cl}$, as    \be 
   S = C \int dt d\tilde x^9 [ \psi \partial_{\tilde x^9} \psi + \cdots ] \sim  C  2\pi \tilde R_9 \int d\tau \psi_{0} \partial_{\tilde x^9} \psi_{0} = \int d\tau  \chi_{cl} \partial_{\tilde x^9} \chi_{cl} ~,~~~~~~ \chi_{cl}  = \sqrt{C 2\pi \tilde R_9 } \psi_0  
   \ee 
   where $\psi_0$ is now the constant mode on the circle.  In the closed string channel, the action of the broken supercharges that generate the multiplet translate into insertions of $\chi_{cl}$. These are the operators $b_I$ and $b^{\dagger I}$ discussed around \nref{SMchi}.  These operators can be replaced by $\psi$ insertions anywhere on the cylinder. 
   However, these differ by normalization factors from similar insertions in the open string channel 
   \be 
    \chi_o =  \sqrt{ \Delta \tau \over 2 \pi \tilde R_9 } \chi_{cl} \ .
   \ee  
  Therefore, we have an identity of the form 
    \be \la{OpCl}
  {\cal Z}_{o}\left({(2J)^4 \over 4!}\right) = Tr\left[ (-1)^F { (2J)^4 \over 4!}e^{ - (2\pi \tilde R_9) H_{open} } \right] =   { ( \Delta \tau)^4  \over ( 2 \pi \tilde R_9) ^4 } \langle B_l | \chi_{cl}^8 e^{ - H_{cl} \Delta \tau } |B_{r} \rangle \ .
  \ee 
   In this discussion, we have suppressed the fermion indices, but there are 8 fermions in the open string channel. In contrast, there are 16 fermions in the closed string channel. However, we have only 8 insertions. This is the necessary number of insertions to amount to 8 supercharge insertions, which is what we have when we look at the 8 fermionic delta functions in \nref{TwoF}. 
   
 Now, let us evaluate the overlap in the right hand side of \nref{OpCl}. The action for the $U(1)$ center of mass degree of freedom in the closed string channel, from \nref{MStr}, involves a  term 
 \be 
S = { N \over R_-} \int dt \half \left({ dx \over d t } \right)^2 
\ee 
where $x$ is the center of mass position. This leads to an energy $H_{cl} = \half { p^2 R_- \over N } $. 
We then get  
\be \la{BDyOvl} \langle B_l | \chi_{cl}^8 e^{ - H_{cl} \Delta \tau } |B_{r} \rangle = {\cal N}^2 V_\parallel \int { d^5 p \over (2 \pi)^5 } e^{ - \half p^2  {R_- \over N } \Delta \tau } = {\cal N}^2 V_\parallel { N^{5/2} \over (2\pi R_- \Delta \tau) ^{5/2} } \ . \ee 
Notice that all oscillator modes are cancelling (among bosons and fermions) in this computation. 
Here, $V_\parallel$ is the volume of the 123 directions which arises due to the overlap of two $|0_\parallel\rangle$ states.  
 This infinite factor is also present in the left hand side of \nref{ConC} from the integral over the bosonic center of mass degrees of freedom. In fact, the left hand side \nref{ConC} is 
 \be \la{OpSi}
 {\cal Z}_o\left({(2J)^4\over 4!}\right) = {\cal I}_4 \times V_\parallel \int { d^3 k \over (2 \pi)^3} \exp\left( - \half { k^2 R_- \tilde R_9^2 (2 \pi)^2 \over N \Delta \tau } \right) =  V_\parallel { ( N \Delta \tau)^{3/2} 
 \over ((2\pi)^3 R_- \tilde R_9^2 )^{3/2} } 
 \ee 
 where we used \nref{Ind8} and  that in the open string channel the center of mass motion modes have the action 
 \be 
  S ={ N \Delta \tau \over R_- 2\pi \tilde R_9 } \int d\tilde x_9  \half ( \partial_{\tilde 9} \vec x )^2 \ .
  \ee

Equating the left and right hand sides of \nref{OpCl}, given in \nref{OpSi} and \nref{BDyOvl}, we get 
  \be \la{NorDe}
  V_{\parallel} { ( N \Delta \tau)^{3/2} 
 \over ((2\pi)^3 R_- \tilde R_9^2 )^{3/2} }  = {\cal N}^2 { (\Delta \tau)^4 \over  ( 2 \pi \tilde R_9)^4 }  V_\parallel { N^{5/2} \over (2 \pi R_- \Delta \tau)^{5/2 } } ~,~~~~\longrightarrow ~~~~ \boxed{ { \cal N } = 2 \pi \sqrt{  R_- \tilde R_9 \over N } } \ .
 \ee 
A consistency check is that there is no $\Delta \tau$ dependence in ${\cal N}$. 

 In conclusion, we have used the two point function to find the normalization constant ${\cal N}$ in the boundary state. 
 
 \subsection{The three point functions and the amplitude } \label{sec:threepoint}
 
 We now turn to the three point amplitude. We will consider the configuration in figure \ref{D3Amplitude}(b), where $N_1 + N_2 \to N_3 $. On the left side we have two sub-blocks of sizes $N_1$ and $N_2$ which are widely separated and on the right we have a single block of size $N_3$. 
  
  There are again two pictures of the underlying process. The first picture is the open string picture, which involves computing an index ${\cal I}_{6}$ (up to the center of mass). The six angular momentum insertions soak up twelve  fermionic zero modes, eight from the center of mass motion, or the $U(1)$ part, and four from the breaking of half of the eight supercharges of the $SU(N)$ theory by our configuration.   
  
  The ``closed string channel'' picture is one that can be viewed as involving the amplitude, at least in the large $\Delta \tau $ limit. From now on we consider this limit of large $\Delta \tau =\tau_r -\tau_l$ and large $x_r^8$ (see figure \ref{D3Amplitude}), with $x_r^8/\Delta \tau$ fixed.    The left boundary state can be decomposed into a product of two boundary states, $\langle B_l | = \langle B_1 | \langle B_2|$, for each of the two sub-blocks of size $N_1$ and $N_2$. We   split the euclidean time evolution into three periods: 
  \begin{enumerate}
  \item The initial time interval when the $N_{1}$ and $N_{2}$ blocks are well separated.
  \item The times when $N_{1}$ and $N_{2}$ merge into the $N_{3}$ block.
  \item The late time interval when the $N_{3}$ block propagates by itself.
  \end{enumerate}
   The middle period is centered around some time $\tau_*$ and has some time width that depends on the fundamental constants but it is much smaller than the overall time interval $\Delta \tau$.   
   We then have an expression of the form
  \be \la{3OpCl}
  \begin{split}
  {\cal Z}_o\left({ (2J)^6 \over 6!} \right) &= \langle B_1 | \langle B_2 | { (2J)^6 \over 6! }  \int_{0}^{\Delta \tau} d\tau^{\star} e^{ -\Delta \tau_1 (  H_1 + H_2 )    } {\hat {\bf A }^c }_M e^{ - (\Delta \tau_3 ) H_3 } | B_3 \rangle ~, \\
  & ~~\Delta \tau_1 \equiv \tau_* - \tau_l ~,~~\Delta \tau_3 \equiv \tau_r - \tau_* \ , 
  \end{split}
  \ee 
    where we suppressed the polarization degrees of freedom for notational simplicity. 
   The $J^6$ insertion   will become clearer below. 
 
 In \nref{3OpCl},  ${\hat {\bf A}^c }_M $ is the amplitude without the energy conserving delta function 
  \be \la{AmRec}
  \hat {\cal A}^c_M = \langle p_1 | \langle p_2 |  {\hat {\bf A}^c}_M |p_3 \rangle 2\pi \delta(\sum p_{+}^{i}) ~.
  \ee 
  This results from doing the path integral in the full theory in the middle region with the boundary conditions set by the momenta in the other two regions. The location of this middle region, which we call $\tau_*$ becomes an integration variable. 
 This energy conserving delta function in the full amplitude comes from the integration over $\tau_*$, schematically  
 \be 
 2\pi\delta(\sum p_+^i) ~~\longrightarrow ~~~ \int d\tau_* \ .
 \ee 
In other words, in the index-amplitude relation (\ref{3OpCl}), it is the amplitude stripped of the momentum conserving delta function that appears on the right hand side, not the actual scattering amplitude, because the amplitude event is occurring at a definite time, $\tau_{\star}$. 

  The basic idea is that we will be able to compute the left hand side of \nref{3OpCl}, and then determine the amplitude that appears in the right hand side of \nref{3OpCl}.

  In order to analyze the problem, it is useful to view this configuration as resulting from a configuration for the center of mass $U(1)$ degree of freedom and a configuration for the $SU(N)$ part. On the right side, we have the $SU(N)$ degrees of freedom forming a single bound state, of size $N_3$. The $SU(N_3)$ theory on the right hand side is completely gapped \cite{Witten:1995im}.  On the left side, we have the degrees of freedom forming two separate bound states with sizes $N_1$ and $N_2$. In this case, there is a massless degree of freedom describing the separation between these two blocks. We have the breaking pattern $SU(N_3) \to SU(N_1) \times SU(N_2) \times U(1)$. Each of the $SU(N_i)$ is individually gapped, but the $U(1)$ part gives rise to gapless modes. 
  
  When we think about the left two sub-blocks, we have the action 
  \bea 
   S &=& {1 \over R_- 2 \pi \tilde R_9 } \int d\tau d\tilde x_9 \half \left[  N_1 (\partial x_1)^2 + N_2 (\partial x_2)^2 \right]  
  \cr 
  &=& { 1 \over R_- 2\pi \tilde R_9 } \int d\tau d\tilde x_9 \half \left[ N_3 (\partial x_{cm})^2 +  N_r (\partial x_r)^2 \right]
  \eea
  where we defined 
  \be \la{NrelDe}
  x_1 = x_{cm } + x_r { N_2 \over N_3 } ~,~~~~~~x_2 = x_{cm} - x_r { N_1 \over N_3 } ~,~~~~N_r = {N_1 N_2 \over N_3 } \ .
  \ee 
  Note that $N_r$ is not an integer. Similarly, we can define U(1) fluxes which are   
  \be \la{nrelDe}
  n_{cm} = n_1 + n_2 = n_3 ~,~~~~~~~~n_r = { ( n_1 N_2 - n_2 N_1 ) \over N_3 } \ .
  \ee 
   For the boundary states of blocks 1 and 2, we write 
   \bea 
   \langle B_1 | \langle B_2 | &\sim &  {\cal N}_1  {\cal N} _2  
   \langle 0_{1 \parallel } |\int {d^5 p_1 \over (2 \pi)^5 } \langle p_1 |  \times \langle 0 _{2 \parallel} |\int { d^5 p_2 \over (2 \pi)^5} \langle p_2 | 
   \cr 
   &\sim & {\cal N}_1  {\cal N} _2  
   \langle 0_{cm \parallel }| \int {d^5 p_{cm} \over (2 \pi)^5} \langle p_{cm} |  \times \langle 0 _{r \parallel}| \int { d^5 p_r \over (2\pi)^5} \langle p_r |
   \cr 
   &\sim & \langle B_{cm} | \times \langle B_r| ~,~~~~~{\rm with}~~~~\langle B_r | \equiv {  {\cal N}_1  {\cal N} _2 \over {\cal N}_3 }  
   \langle 0 _{r \parallel}| \int { d^5 p_r \over (2 \pi)^5}\langle p_r |
\la{BrDe} 
\eea
where in the second line the two-body state is re-written in terms of its center of mass and relative momenta, $ p_{cm} = p_1 + p_2  $ and $p_r = (N_2 p_1 - N_1 p_2
)/N_3 $ respectively. Note that the factor of ${\cal N}_3$ comes in when we define $\langle B_{cm} |$.

   The equality \nref{3OpCl} can be separated into two equalities, one for the center of mass, the $U(1)$ part, and one for the relative part, the $SU(N_{3})$ part. In preparation for this splitting, it is useful to write the matrix model amplitude as 
    \bea 
    \hat{\cal A}^c_M  &=& \hat {\cal A}_{Mcm}^c \hat {\cal A}^c_{Mr} 
    \\ 
   \hat {\cal A}_{Mcm}^c &=& (2\pi)^8\delta^8(\sum \vec p^{\, i} ) \delta^4( \bar \eta^{cm}_J + \breta^3_{J})\delta^4(   \eta^{cm K} - \eta^{3K}) \la{Acm} 
\\ \la{Arel} 
\hat {\cal A}_{Mr}^c &=&2\pi  \delta(\sum p^i_+) \delta^4( \bar \eta^{r}_J  ) A(p_r) 
\eea
where $A(p_r)$ is 
an off-shell amplitude that is a function of $p_{r}$, $N_i$ and $n_i$. The delta function for energy conservation fixes the value of $p_r$ in terms of 
  $N_i$ and $n_i$, but we leave it general now to make the derivation below a bit more clear. 
   The goal is to determine the on-shell value of $A$. 
Finally, note that the dependence on the Grassmann variables is fixed by supersymmetry from the discussion around \nref{DelCon}; our choice of ``vacuum'' is the state produced by the M2 branes. This state is a combination of the graviton and the three form potential that is invariant under the $SO(3)\times SO(6)$ subgroup of $SO(9)$ that leaves the M2 brane invariant (that preserves the 123 subspace). The $SO(9)$ is the symmetry around the null momentum of the particle.

The part involving $\hat {\cal A}_{Mcm}^c$ in \nref{Acm} is the same as what appeared in the two point function. This part involves only the $U(1)$ degrees of freedom and the discussion is exactly the same as in the two point function. We subsequently focus on the $SU(N)$ problem, which schematically involves 
\be \la{SUNOpCl}
 \widetilde{ {\cal Z}}_{o}\left({ (2 J)^2 \over 2} \right) \sim \int d\tau^{\star} \langle B_r | {(2J)^2\over 2} e^{ - \Delta \tau_1  H }  \hat {\bf A}_{Mr}^c |  0 \rangle 
 \ee 
  where the $|0\rangle $ on the right hand side is the $SU(N_3)$ ground state. The tilde means we are considering the $SU(N)$ problem.  
 
 We  now analyze this problem in more detail. This is a completely self contained problem in the 1+1 dimensional $SU(N)$ gauge theory on an interval. On the right side, there is an $SU(N_3)$ gapped ground state, as long as an appropriate non-zero value of $n_3$ is chosen such that it is coprime with $N_3$.   The details of the D3 brane on the right side do not matter much. They just give a constant that was already included in the normalization factor ${\cal N}_3$. On the left side, the $SU(N_3)$ is broken to $SU(N_1) \times SU(N_2) \times U(1)_r$ by the boundary conditions. These boundary conditions involve the statement that the relative flux in the $U(1)_r$ part to be $n_r$ and the $8^{th} $ position to be $x_r =x^8_r$, see figure \ref{D3Amplitude}. 
  We see that the vacuum expectation value of $X^8_r$ is non-zero near the boundary and then it becomes zero at some intermediate value of Euclidean time   $\tau_*$. This is just a classical picture. In the full quantum picture, there can be fluctuations around this configuration. 
  The presence of a non-zero gradient for $X^8_r$ as well as the flux $n_r$ breaks  4 the 8 supercharges that are preserved by the boundary condition. This leads to four fermion zero modes that are soaked up by the insertion of $J^2$ in \nref{SUNOpCl}.   
  
  Notice that, in contrast with the two point function case, the relative fields all vanish on the right side. This implies that there is no volume factor and that 
  \be \la{ZopIn}
  \tilde {\cal Z}_o\left({(2J)^2\over2} \right) = \pm ( n_1 N_2 - n_2 N_1) 
  \ee 
  This follows from a logic very similar to that in Sen's computation \cite{Sen:2012hv}. It is identical to that computation if we are willing to consider the index in the full string theory. On the other hand, if we interpret the computation in \cite{Sen:2012hv} as a computation of dyons in ${\cal N}=4$ SYM, then our problem is slightly different because we are considering the problem from the point of view of the $1+1$ field theory. These two problems are connected by the Nahm transform \cite{Nahm:1979yw,Diaconescu:1996rk}, so we expect that we can phrase Sen's computation of the index as a computation in the 1+1 dimensional field theory. 
 
 Let us now turn to the right hand side of \nref{SUNOpCl}. The insertion of $J^2$ translates into the insertion of four fermions that are canonically normalized in the open string channel, $\chi_o$. It is convenient to translate these to fermion zero modes canonically normalized in the closed string channel 
 \be 
 {\chi}_o = \sqrt{ \Delta \tau_* \over 2\pi \tilde R_9 } \chi_{cl} 
 ~,~~~~~~{ (2J)^2 \over 2 } \sim \chi_o^4 \sim { (\Delta \tau_*)^2 \over (2\pi \tilde R_9)^2 } \chi_{cl}^4 ~,~~~~~\Delta \tau_*   = \tau_* - \tau_l  \la{ClSu}
 \ee 
 where $\Delta \tau_* $  is the Euclidean time between the left boundary and the point where the vev becomes zero. The location of this point is integrated over as we discuss below in more detail. The insertion of $\chi_{cl}$  then corresponds to the action of the supercharges and is related to the four Grassmann delta functions in \nref{Arel}.

 Now, let us discuss the contributions from the bosonic momentum eigenstates in the boundary state. 
 On the left side, we have $\langle B_r| $ as in \nref{BrDe}. 
 On the right side, we have a state with $X_r=0$, so we expect that we are simply expanding a delta function that contains all momenta. Along the longitudinal directions (directions 1,2,3), the state $|0\rangle_\parallel $ selects the zero momentum part. For the other five dimensions (directions 4,5,6,7,8), we still need to do an integral, which is of the form 
\be \la{MomCon}
 \int {d^5 p_r \over (2\pi)^5} e^{i p_r . x_r }  \exp( - \half { p_r^2 R_- \over N_r } \Delta \tau_* )A(p_{r})  =  \left( {N_r \over 2\pi R_- \Delta \tau_* } \right)^{5/2}   \exp\left( - \half { x_r^2 N_r \over R_- \Delta \tau_* } \right) A(p_{r}^{s})
 \ee 
  where $p_{r}^{s}$ is the saddle point for $p_{r}$, namely $p_{8r}^s = i {  N_r\over R_-} {x_r^8 \over \Delta \tau_* }$. 
    The term involving $x_r$ in \nref{MomCon} arises because we are fixing the boundary value of $x_r=x_r^8$ to a non-zero value at the left boundary, which involves a displacement in the $X^8$ coordinate, see figure \ref{D3Amplitude}. We can interpret the exponential in \nref{MomCon} as the energy due to the gradient in $X^8$. There is an extra contribution to the energy from the electric flux, which leads to an extra factor of the form 
  \be \la{FluxE}
  \exp\left( - \half{  n_r^2  R^3_- \Delta \tau_*   \tilde R_9^2 \over N_r l_p^6 } \right) \ .
  \ee 
  Putting together all the $\tau_*$ dependent terms from \nref{ClSu} \nref{MomCon} \nref{FluxE} and integrating over $\tau_* $, we find a integral amiable to a saddle point approximation:
  \be \la{TauInt}
  \int_0^\infty  d(\Delta \tau_*) { 1 \over (\Delta \tau_* )^{1/2}  }  \exp\left ( - \half { a \over \Delta \tau_* } -\half  b \Delta \tau_* \right)f(\Delta \tau_*)= \sqrt{ 2 \pi \over b }    \exp( - \sqrt{ab} )  f(\Delta \tau_*^s) 
   \ee 
   where $\Delta \tau_*^s$ is the saddle point of the $\Delta \tau_*$ integral.\footnote{Note that $A(p_{r}^{s})$ depends on $\tau_{\star}$ because $p_{r}^{s}$ depends on $\tau_{\star}$.} Putting all of this together, the right hand side of \nref{SUNOpCl} is then 
   \be \la{ClCha}
  \langle B_r | {(2J)^2\over 2 } e^{ - \Delta \tau_1  H } \hat {\bf   A}_{Mr}^c |  0 \rangle  = {{ \cal N}_1 {\cal N}_2 \over {\cal N}_3 }{ 1 \over (2\pi)^2 \tilde R_9^2} \left({ N_r \over R_-} \right)^{5/2} {1 \over (2 \pi)^2} { N_r^{\half} l_p^3\over \tilde R_9 n_r  R_-^{3/2} }  \exp\left( -x_r { n_r \over R_9 } \right)  A(p_{r}^{s})
   \ee 
   where $A(p_{r}^{s})$ was defined in \nref{Arel}.  The saddle point approximations in (\ref{MomCon}) and (\ref{TauInt}) together localize $p_{r}^{s}$ to its on-shell value 
   \be 
   p_{8r}^s = i  { n_r \over R_9 } = { N_2 p^1_8 - N_1 p^2_8 \over N_3} \ .
   \ee 
    The remaining exponential in \nref{ClCha} could be viewed as the energy contribution at the saddle point of the $\tau_*$ integral \nref{TauInt}. This is just the classical energy of the configuration in figure \ref{D3Amplitude}.   Since the configuration is BPS, we expect that its energy should be equal to the appropriate ``central charge'' of the supersymmetry algebra. This central charge can depend on the boundary conditions. In fact, we see that the exponential in \nref{ClCha} depends in a simple way on the boundary value $x_r$ and the electric field flux $n_{r}$. We therefore argue that the exponential term in \nref{ClCha} should be removed once we subtract the central charge and will drop it.

    Using \nref{NrelDe}, \nref{nrelDe}, and \nref{NorDe},  and equating \nref{ClCha} to \nref{ZopIn} we get 
   \bea 
   \tilde {\cal Z}_o\left({(2J)^2\over2 }\right) &= & \langle B_r | {(2 J)^2\over 2 } e^{ - \Delta \tau_1  H } \hat {\bf  A}_{Mr}^c |  0 \rangle 
   \cr 
    ( n_1 N_2 - n_2 N_1) &= & { 1 \over (2\pi)^3} \sqrt{N_3 \over N_1 N_2 } \sqrt{R_- \tilde R_9} { 1 \over \tilde R_9^2} \left({ N_r \over R_-} \right)^{5/2} { N_r^{\half} l_p^3\over \tilde R_9 n_r  R_-^{3/2} }   A(p_{r}^{s})
   \eea
   which then leads to 
  \be \label{finalres}
  \boxed{  A(p_{r}^{s}) =(2\pi)^3  { l_p^{9/2} R_-  \over R_9^{5/2} }{ 1 \over \sqrt{N_1 N_2 N_3 } } { ( n_1 N_2 - n_2 N_1)^2 N_3^2 \over N_1^2 N_2^2 } } \ .
   \ee 
   Equation \nref{finalres} agrees precisely with \nref{ThreeF}.  One might worry that \nref{finalres} is only approximate because we are using perturbative techniques in the initial time interval even though we are claiming the result is valid for generic values of $N$ and $n$. In principle, the integrals over $p_{r}$ and $\tau_{\star}$ could be corrected by fluctuations around the saddle point, where the precise corrections are fixed by $N$ and $n$. However, the actual widths of the $p_{r}$ and $\tau_{\star}$ saddle points can be taken to be arbitrarily small by taking $x_{r}, ~\Delta \tau$ to be arbitrarily large.  Therefore, the perturbative calculation for the initial time region is exact in the large $x_{r}, ~\Delta \tau $ limit.     
   
   In conclusion, the amplitude computed using the matrix model coincides with the amplitude expected in gravity. Having obtained the amplitude when $R_9$ is finite, we can also take the limit that $R_9 \to \infty $ and $n^i\to \infty $ keeping $p^i_9 = n_i/R_9$ fixed ($p_8$ also remains fixed). In this limit, we recover the three point amplitude in the 0+1 dimensional matrix model. We can further take the limit $N_i\to \infty$, keeping $p^i_- = { N_i/R_-} $ fixed. This gives the amplitude in the non-compact eleven dimensions. Note that the amplitude agrees precisely even at finite $N$ and $n$. For the BFSS conjecture, this agreement was not necessary, only the large $N$ limit had to agree with gravity.  Perhaps there is some argument, beyond this explicit computation,  that explains why the amplitudes had to agree even at finite $N$ and $n$.

    Notice that in the weakly coupled IIA limit, $R_9 \to 0$,  these amplitudes do {\it not} go over to the fundamental string amplitudes when $n_i$ are non-zero. They are amplitudes that involve scattering of bound states of D0 branes in the IIA theory.  It is amusing that in the large $N$ limit, these amplitudes become very similar to the problem of scattering D0 branes in the original BFSS matrix model.

 In appendix \ref{Nzero}, we discuss the three point amplitude in the special case that one of the gravitons has zero longitudinal momentum, $N=0$. 
 
    \section{Conclusions } 

In this paper,  we have computed the three graviton amplitude in the BFSS model. Crucially, all three particles have non-zero longitudinal momentum, or $N_i > 0$, so this computation probes strongly coupled regime of the matrix model. We have shown that the amplitude agrees with the expected form in M-theory. To do the computation, we used that the three point boundary conditions preserve some  supersymmetries and related  the amplitude to an index, a protected quantity, which we in turn used to fix the amplitude, including its overall numerical coefficient. The fact that certain supersymmetric amplitudes can be computed by relating them to other problems has been used before many times, e.g. \cite{Antoniadis:1993ze,Alday:2007he}. To make the identification, we found it useful to view the amplitude as arising from the exchange of gravitons sourced by M2 brane ``operators.'' For the case of the two point function, a  somewhat similar discussion was contained in \cite{Polchinski:1997pz}. In this paper, we considered the case where the gravitons interact as they go between different brane sources. 

This three point amplitude  can be used to derive the soft theorems \cite{Weinberg:1965nx,Cachazo:2014fwa}, as we show in a separate publication \cite{Herderschee:2023bnc}. 

There are a number of future directions. 

It is natural to wonder whether other supersymmetric amplitudes could be computed by using a similar method. For example, can other string network indices be used to compute higher point amplitudes? Particularly interesting amplitudes are those related to higher curvature corrections of the gravitational action. In addition, string networks like the ones we considered sometimes have moduli. For example, in the network we considered, we could add a small triangle at the interaction vertex so that the three individual three point amplitudes are sewn together. For the particular case of the three point amplitudes, such configurations cannot become very large because of the polarizations. However, for higher point amplitudes we would need to worry about such configurations, see \cite{Sen:1997xi,Sen:2012hv}. 

We could also consider three point amplitudes in different theories. For example, the graviton three point function in heterotic string theory has an $\alpha'$ correction. It would be interesting to see whether this is reproduced by a similar computation in the corresponding  matrix model \cite{Banks:1997it,Motl:1997tb}.  Here the three point function has a correction that depends on a continuous parameter $(\alpha')$, which seems to be a problem for the strategy pursued in this paper. 

Since the three point amplitude is an essential element in the BCFW \cite{Britto:2005fq} construction of gravity amplitudes, one could wonder whether a similar argument could show that all gravity amplitudes are reproduced by the matrix model. Of course, we expect this to work only for the low energy gravity amplitudes rather than the full M-theory amplitudes which would have a more complicated analytic structure as a function of momenta. One significant hurdle is arguing that the contour at infinity in the BCFW derivation should vanish in the supergravity regime. 

Finally, it might also be possible to use the supersymmetric localization methods developed in \cite{Asano:2014eca,Asano:2012zt} for the plane wave matrix model \cite{Berenstein:2002jq} to compute amplitudes. It seems that the amplitude computation in that model might be related to a tunneling computation between two different vacua of the model, but we did not manage to do the computation in that way.

\subsection*{Acknowledgments}

We   thank O. Aharony, T. Banks, M. Dine, Y-Z. Li,   N. Miller, S. Mizera, N. Seiberg, A. Strominger, Z. Sun, A. Tropper, N. Valdes-Meller,    H. Verlinde, and T. Wang for discussions.   

J.M. is supported in part by U.S. Department of Energy grant DE-SC0009988. A.H. is supported by the Simons Foundation.  

\appendix

\section{Computation of the three point amplitude when one particle has no light cone momentum} 
\la{Nzero} 

\begin{figure}[h]
    \begin{center}
    \includegraphics[scale=.4]{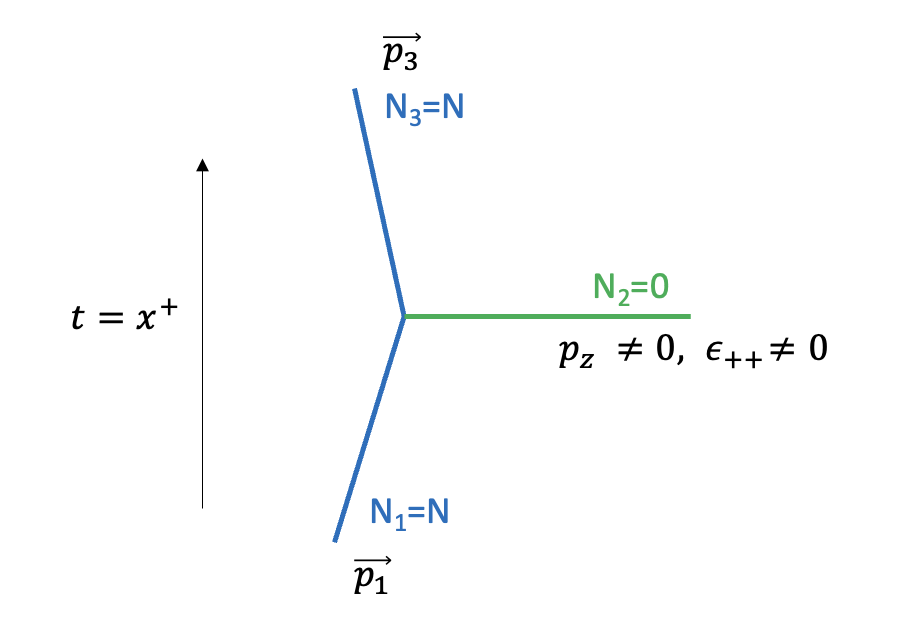}
    \end{center}
    \caption{The scattering configuration for the three point function where one of the particles has zero momentum. It leads to the insertion of the operator \nref{PWOP} in the matrix   quantum mechanics.     }
    \label{NzeroFig}
\end{figure}

In this appendix, we discuss the computation of the three point function in the special case when $N_2=0$, $N_1 = N_3 = N$, see figure \ref{NzeroFig}.  
In this case,  the gravity mode with $N_2 =0$ can be viewed simply as a graviton  with some momentum $p_z\not =0, ~p_{\bar z } =0$ in two of the transverse dimensions. We can also choose the polarization vector $ \epsilon_{++} \not =0$ and the rest of the components equal to zero. 
Then, we simply have the original D0 brane action interacting with a background gravitational plane wave metric of the form 
\be 
ds^2 =  -2 dx^+ dx^- + 2 d z d \bar z + d \vec y^2 +  (dx^+)^2 \epsilon_{++} e^{ i p_z z } 
\ee 
This induces and extra term in the Matrix model Lagrangian \nref{LagBFSS} that is simply a potential term of the form 
\be \la{PWOP}
\int dt \epsilon_{++} \textrm{Tr}[ e^{ i p_z Z } ] 
\ee

The computation of the amplitude involves expanding the path integral to leading order in $\epsilon_{++}$ and this results to an insertion of \nref{PWOP} as an operator. In other words, we need to compute  
\be 
\langle N, \vec p_1 | \int dt \epsilon_{++} \textrm{Tr}[ e^{ i p_z Z } ] |N, \vec p_3 \rangle 
\ee 
The computation of this expression involves the $U(1)$ degrees of freedom, which simply give rise to the 9 spatial momentum delta functions. The integral over $dt$ produces the energy conservation delta function.  
This $U(1)$ part reproduces the amplitude because it is simply the superparticle action in lightcone gauge interacting with a plane wave, which is merely one way in which we could compute the amplitude in gravity. Notice that this also implies that the overall coefficient also reproduces gravity. 

What remains is to compute 
\be \la{ExpSU}
{ 1 \over N } \langle 0| \textrm{Tr}[ \exp(  i p_z Z )] | 0 \rangle 
\ee 
in the $SU(N)$ theory. Here $|0\rangle$ is the bound state. We do not know it explicitly. However, we know it is $SO(9)$ invariant. On the other hand, the $Z $ that appears in the exponent carries $SO(2)$ charge. So, if we expand the exponential, all the nonzero powers of $Z$ have a zero expectation value by the symmetries of the problem.  Therefore, the full expectation value \nref{ExpSU} is one. 
One might be worried that this might be somewhat ill defined for very high powers of $Z$ because the bound state wavefunction decays only as a power law in the flat directions. In fact, we can give a more solid argument by introducing a small mass term $\mu$, so that we consider a computation in the plane wave matrix model \cite{Berenstein:2002jq}. This computation can be performed using the localization method in  \cite{Asano:2014eca,Asano:2012zt}, after identifying the bound state with the maximal $SU(2)$ representation in the discussion in  \cite{Asano:2014eca,Asano:2012zt}. 

\bibliographystyle{apsrev4-1long}
\bibliography{GeneralBibliography.bib}
\end{document}